\documentclass[12pt]{iopart}

\usepackage{epsfig}
\usepackage{color}
\newcommand{\ful}{\mbox{C$_{\mbox{\scriptsize{60}}}$}}
\newcommand{\vscf}{\mbox{$\delta V$}}
\newcommand{\eq}[1]{Eq.~\ref{#1}}

\begin{document}
\title[First prediction of C$_{60}$-to-confined-atom backward RICD]
{First prediction of inter-Coulombic decay of C$_{60}$ inner vacancies through the continuum of confined atoms}

\author{Ruma De$^1$, Maia Magrakvelidze$^{1,*}$, Mohamed E Madjet$^2$, Steven T Manson$^3$, and Himadri S Chakraborty$^{1}$}

\address{$^1$D. L. Hubbard Center for Innovation and Entrepreneurship, Department of Natural Sciences, Northwest Missouri 
State University, Maryville, Missouri 64468, USA}
\address{$^2$Qatar Environment and Energy Research Institute, Hamad Bin Khalifa University, Qatar Foundation, P.O. Box 5825, Doha, Qatar}
\address{$^3$Department of Physics and Astronomy, Georgia State University, Atlanta, Georgia, USA}
\address{$^*$Current address: Department of Physics, Kansas State University, Manhattan, Kansas 66502, USA}

\ead{himadri@nwmissouri.edu}

\begin{abstract}
Considering the photoionization of Ar@$\ful$ and Kr@$\ful$ endofullerenes, the decay of $\ful$ innershell excitations through the outershell continuum of the confined atom {\em via} the inter-Coulombic decay (ICD) pathway is detailed. Excitations to atom-$\ful$ hybrid states demonstrate coherence between ICD and electron-transfer mediated decay (ETMD). This should be the dominant above-threshold decay process for a variety of confined systems, and the strength of these resonances is such that they should be amenable for study by photoelectron spectroscopy.

\end{abstract}

\newpage

The resonant energy transfer from a donor to a near-neighbor acceptor molecule is a ubiquitous phenomenon in matter~\cite{scholes2003review}. In complex materials, comprising a variety of light absorbing components called chromophores with clean absorption and fluorescence bands, the photoexcitation at the donor site is commonly followed by a migration of the energy to a closely separated acceptor chromophore which can subsequently relax to emit a fluorescent photon. The Coulomb interaction between donor-acceptor active electrons is the conduit of such non-local energy transfer. This fluorescence resonant energy transfer (FRET) spectroscopy has a vast range of applications in biological, nanoscale~\cite{beljonne2009bio-nano}, and nano-bio hybrid~\cite{medintz2009bio-nano-hyb} systems. For relatively smaller aggregations of matter, such as clusters and dimers with limited electronic and vibrational degrees of freedom, the excited state of an inner electron at the donor site can often be embedded in the ionization continuum of the acceptor, resulting in a non-radiative release of an acceptor electron as the excitation energy relocates. This phenomenon, known as inter-Coulombic decay (ICD), was predicted~\cite{cederbaum1997firstTh} and was first observed for Ne clusters~\cite{marburger2003firstExp} some years ago. Novel experiments on the ICD process focusing on fundamental science are abundant~\cite{jahnke2004rareDimer,oehrwall2004rareClust,grieves2011surface,jahnke2010water1,mucke2010water2}. Furthermore, ICD driven slow electrons may potentially find applications in controlled radiation damage in medical sciences~\cite{gokhberg2014MedApp,trinter2014MedApp}. On the other hand, if the original excitation entails the transit of an electron from one center to the excited state of another center, then its decay with the creation of an outer shell vacancy on either center is called resonant electron-transfer mediated decay (ETMD)~\cite{foerstel2011etmd-expArKrCluster,maza2011etmd-expArDimer}. 

From a fundamental spectroscopic viewpoint, probing resonant ICD (RICD) processes~\cite{barth2005ricdExp1,aoto2006ricdExp2,kim2013DimExp} in relatively simpler gas-phase materials is of great interest~\cite{barth2005ricdExp1,aoto2006ricdExp2,trinter13HeNe,najjari10TwoCenter}. One class of such systems undergoing significant current theoretical and experimental scrutiny is atomic endofullerene complexes, in which an atom is encapsulated in a fullerene molecule. These are unique heterogeneous and nested dimers of weak atom-fullerene bonding exhibiting very different electronic properties from the atom to the fullerene. From the experimental side, the synthesis techniques for these materials are fast-developing~\cite{popov2013endoSynth} with the huge advantage of their room-temperature stability. Moreover, these materials promise a plethora of applied contexts~\cite{popov2013endoSynth}. The earliest prediction of the shortening of the decay time of a Ne excited state through the ICD pathway when Ne is confined in $\ful$ was made in 2006~\cite{averbukh2006endo-icdTh}. Later, such broadening of atomic Auger lines due to non-local Coulomb-mediated decay in endofullerenes were also suggested by others~\cite{korol2011,amusia2006}. The first detailed calculations of RICD resonances corresponding to Ar inner $3s$ excitations in the photoionization of the levels of the encapsulating $\ful$, the atom-to-fullerene {\em forward} RICD, was recently performed by us~\cite{javani2014-rhaicd}. In addition, a dominant and novel effect was found in the coherence between the Auger and ICD transition amplitudes to produce resonance structures in the photoionization of atom-fullerene hybridized states~\cite{javani2014-rhaicd,magrakvelidze2015-rhaicd}. These resonant hybrid Auger-ICD (RHA-ICD) features, with their various shapes and widths, bear signatures of this coherence. However, to the best of our knowledge, no RICD of a $\ful$ inner vacancy producing a purely atomic outer vacancy in the encaged atom, the fullerene-to-atom {\em backward} RICD, has been predicted until now. Only, as a reverse analogy, but for a non-resonant spectral effect from atom to fullerene, even when the participating charge densities are pure non-hybrid, was seen earlier~\cite{mccune2010-atom-to-c60}.

In this communication, considering the photoionization of the valence $ns$ subshell of the atoms $X$, Ar and Kr, in $X$@$\ful$, we predict autoionizing resonances from the RICD of $\ful$ inner excitations. Owing to the hybridization of some of the excited states of the compound, the RICD amplitude is also found to admix coherently with the ETMD process. The results, along with our previous findings~\cite{javani2014-rhaicd,magrakvelidze2015-rhaicd}, complete the full ICD landscape in a photon-driven endofullene molecule, highlighting these materials, in gas or condensed phase, as possible candidates for experiments. 

Kohn-Sham density functional theory is used to describe the ground state electronic structure of the compounds~\cite{madjet2010xeFull}. The $\ful$ molecule is modeled by smudging sixty C$^{4+}$ ions over a classical spherical jellium shell, fixed in space, with an experimentally known $\ful$ mean radius 3.5 \AA and thickness $\Delta$. The nucleus of the confined atom is placed at the center of the sphere. The Kohn-Sham equations for the system of a total of $240+N$ electrons ($N=18$ for Ar, $N=36$ for Kr and 240 delocalized electrons from $\ful$) are then solved to obtain the electronic ground state properties in the local density approximation (LDA). The gradient-corrected Leeuwen and Baerends exchange-correlation (XC) functional [LB94]~\cite{van1994exchange} is used for the accurate asymptotic behavior of the ground state radial potential
\begin{equation}\label{lda-pot}
V_{\scriptsize \mbox{LDA}}(\mathbf{r}) = V_{\mbox{\scriptsize jel}}({\bf r}) - \frac{z_{\mbox{\tiny atom}}}{r} + \int d\mathbf{r}'\frac{\rho(\mathbf{r}')}{|\mathbf{r}-\mathbf{r}'|} + V_{\scriptsize \mbox{\scriptsize XC}}[\rho(\mathbf{r})],
\end{equation} 
which is solved self-consistently in a mean-field framework. The requirement of charge neutrality produced $\Delta =$ 1.3 \AA, in agreement with the value inferred from experiment~\cite{ruedel2002oscExp}.

The time-dependent local density approximation (TDLDA) is employed to simulate the dynamical response of $\ful$ to incident photons~\cite{madjet-jpb-08}. The dipole operator, $z$, corresponding to light that is linearly polarized in $z$-direction, induces a frequency-dependent complex change in the electron density arising from dynamical electron correlations. This can be written, using the independent particle (IP) susceptibility $\chi_0$, as
\begin{equation}\label{ind_den2}
\delta\rho({\bf r};\omega)={\int \chi_0({\bf r},{\bf r}';\omega) 
                           [z' + \vscf({\bf r}';\omega)] d{\bf r}'},
\end{equation}
in which
\begin{equation}\label{v_scf}
\vscf({\bf r};\omega) = \int\frac{\delta\rho({\bf r}';\omega)} {\left|{\bf r}-{\bf r}'\right|}d{\bf r}'
     +\left[\frac{\partial V_{\mbox{xc}}}{\partial \rho}\right]_{\rho=\rho_{0}} \!\!\!\!\delta\rho({\bf r};\omega),
\end{equation}
where the first and second terms on the right hand side are, respectively, the induced changes of the Coulomb and the exchange-correlation potentials. Obviously, $\vscf$ includes the dynamical field produced by important electron correlations within the linear response regime. In this method, the photoionization cross section corresponding to a bound-to-continuum dipole transition $n\ell\rightarrow k\ell^\prime$ is given by
\begin{equation}\label{cross-pi}
\sigma_{n\ell\rightarrow k\ell'} \sim |\langle k\ell'|z+\vscf|n\ell\rangle|^2,
\end{equation}
where the TDLDA matrix element ${\cal M} = {\cal D} + \langle\vscf\rangle$, with ${\cal D}$ being the independent-particle LDA matrix element.

\begin{figure}
	\centerline{\psfig{figure=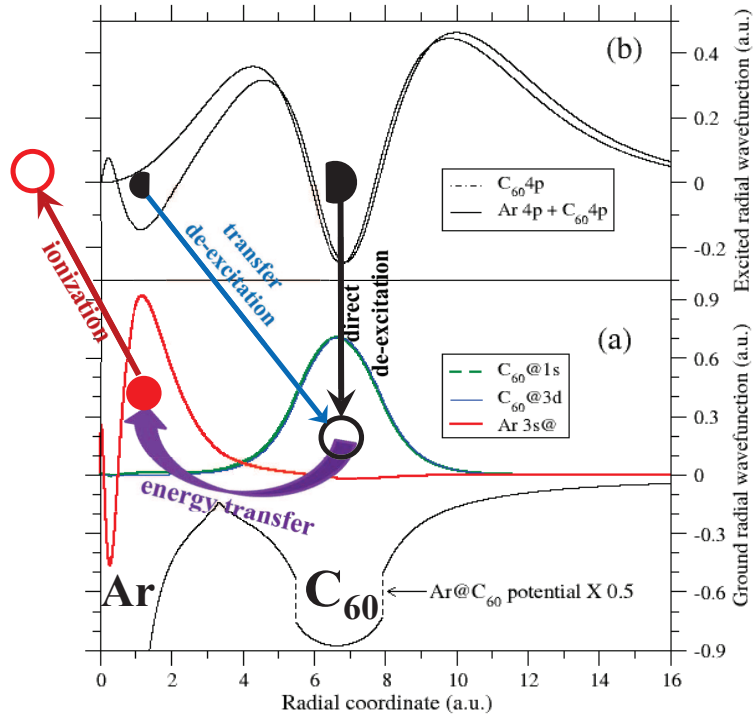,height=10.0cm,width=10.5cm,angle=0}}
	\caption{(Color online) (a) Ground state radial wavefunctions of the Ar@$\ful$ complex: these are identified as pure Ar $3s$ and pure $\ful$ inner $1s$, $3d$. The scaled radial potential of the system is also shown. (b) The radial wavefunction of an empty $p$ state of the complex that can be excited by dipole transitions from $\ful$ states in panel (a). This state is hybridized between Ar $4p$ and $\ful$ $4p$; the wavefunction of the later is also shown. The coherence of ICD and ETMD amplitudes in the emission of Ar $3s$@ photoelectrons is schematically illustrated.}
	\label{fig:xe@c60-figure1}
	\vspace{-0.35cm}
\end{figure}

It is well-known that in $X$@$\ful$ the atomic valence $np$ electrons strongly hybridize with the energetically shallower $p$ electrons of the host $\ful$ \cite{javani2014-rhaicd,magrakvelidze2015-rhaicd,morscher2010strong}. The subvalent $ns$ levels of $X$, however, maintain their purity, as seen in Figure 1(a) which shows predominantly atom-like Ar $3s$@ radial wavefunction from the ground LDA spectrum of Ar@$\ful$. Single-electron excitations from a number of $\ful$ inner levels $n'\ell$ (whose ionization thresholds are indicated in Figure 2) occur at energies higher than the Ar $3s$@ and Kr $4s$@ thresholds, 30.1 eV and 26.5 eV respectively. Of these $\ful$ levels, @$3d$ and @$1s$ wavefunctions are presented in Fig.\,1(a). (Although these levels are the quantum states of the whole compound, we use $n\ell$@ and @$n\ell$ respectively to ascertain their atom- or $\ful$-dominant character.) Using the well-known approach by Fano~\cite{fano1961} to describe the dynamical correlation through the interchannel coupling, the RICD amplitudes of these $\ful$ photo-vacancies {\em via} $X$ $ns$@ ionization can be expressed by $M^{\mbox{\scriptsize d-c}}$ that denotes the coupling of $\ful$ discrete (d) excitation channels $@n'l\rightarrow\eta\lambda$ with the $ns@\rightarrow kp$ continuum (c) channel of $X$. Following~\cite{javani2014-rhaicd}, $M^{\mbox{\scriptsize d-c}}$ can be written as:
\begin{eqnarray}\label{dc-mat-element}
 {M}^{\mbox{\scriptsize d-c}}_{ns@\rightarrow kp}(E) &=& \displaystyle\sum_{@n'\ell} \sum_{\eta\lambda} \frac{\langle\psi_{@n'\ell\rightarrow\eta\lambda}|\frac{1}{|{\bf r}_1-{\bf r}_2|}
|\psi_{ns@\rightarrow kp}(E)\rangle}{E-E_{@n'\ell\rightarrow\eta\lambda}} {\cal D}_{@n'\ell\rightarrow\eta\lambda},
\end{eqnarray}
where $E_{@n'\ell\rightarrow\eta\lambda}$ and ${\cal D}_{@n'\ell\rightarrow\eta\lambda}$ are, respectively, excitation energies and LDA matrix elements of channels $@n'\ell\rightarrow\eta\lambda$ and $E$ is the photon energy corresponding to the $ns@\rightarrow kp$ transition. In \eq{dc-mat-element} the $\psi$ are independent-particle (LDA) wavefunctions that represent the final states (channels) for transitions to excited/continuum states. Obviously, the Coulomb matrix element in the numerator of \eq{dc-mat-element} acts as the ``corridor'' for energy transfer from the $\ful$ de-excitation across to the atomic ionization process, producing ICD resonances in the @$ns$ cross sections.

\begin{figure}
\vskip 0.5 cm
	\centerline{\psfig{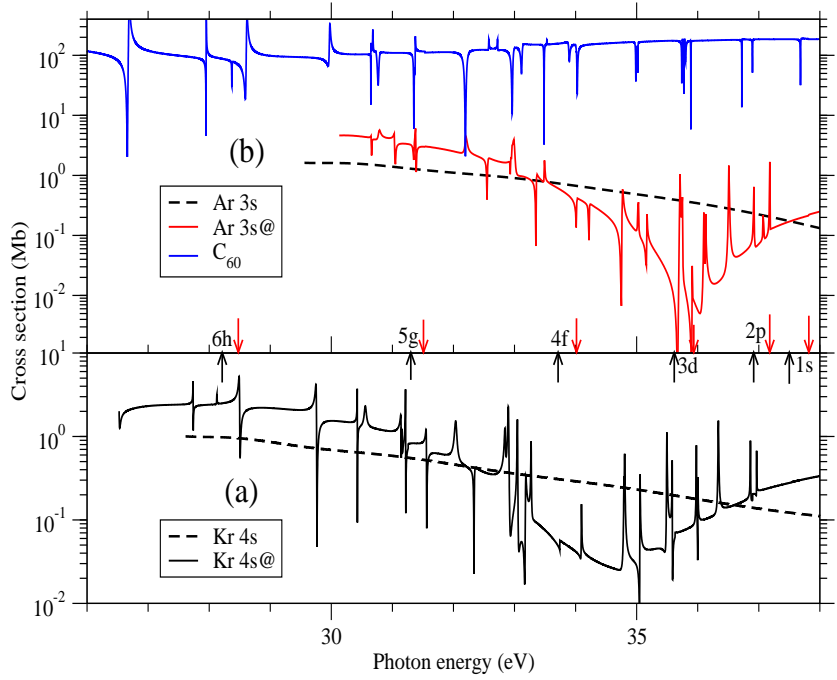}}
	\caption{(Color online) Photoionization cross sections of $4s$ subshell for free and confined Kr (a) and $3s$ subshell of free and confined Ar (b). The total cross section of empty $\ful$ is also presented. Various $\ful$ ionization thresholds of respective complexes are shown.}
	\label{fig:ar@c60-figure2}
	\vspace{-0.1cm}
\end{figure}

These $\ful$-to-$X$ ICD resonances are displayed in Figure 2 both for $4s$ [Fig.\,2(a)] and $3s$ [Fig.\,2(b)] photoionization of, respectively, confined Kr and Ar. As seen, the spectra are rather dramatically structured. Note that the corresponding results for free atoms are flat, since the current energy range does not include any regular autoionizing (Auger) decay of atomic innershell vacancies. The resonances in Figs.\,2 are strong, of varied shapes, and should be easily accessible {\em via} photoelectron spectroscopy. Furthermore, the narrow width of these resonances, which is very different than characteristic atomic Auger resonances that are generally broad, is directly related to the $\ful$ excitations. Indeed, like generic cluster wavefunctions,  $\ful$ wavefunctions are typically delocalized. spreading out over a large volume, in contrast to atomic localized electrons (see, Fig.\,1). Since the autionization rate involves the matrix element of $1/r_{12}$ [\eq{dc-mat-element}], spread-out (delocalized) wavefunctions translate to a decrease in the value of the matrix element, as compared to atomic compact (localized) wavefunctions. 

Fig.\,2 exhibits three particularly notable features: (i) The above-threshold vacancy decay is completely dominated by ICD. (ii) For both Kr $4s$@ and Ar $3s$@, the characteristic Cooper minima are moved lower in energy to 35 eV and 36 eV from their well known positions of 41 eV and 42 eV~\cite{magrakvelidze-ar-delay-2015}, respectively, for free atoms. This shift is a consequence of the atom-$\ful$ dynamical coupling, particularly the coupling of $X$$ns@\rightarrow kp$ ionization with a host of $\ful$ continuum channels. Note that this coupling was not included in \eq{dc-mat-element}, which only captures the resonances, but is certainly present in the full dipole matrix element ${\cal M}$ (see \eq{cross-pi}). (iii) A comparison with the empty $\ful$ cross section in Fig.\,2(b) reveals a few extra resonances in both the Kr $4s$@ and the Ar @$3s$ results. These are present owing to the additional excited states in the excited spectrum of the whole compound, since it now also includes the excited states of the caged atom. In fact, many of these excited states of the compound must be hybrids between the atomic and $\ful$ pure states -- a fact addressed below suggests that some of these resonances are the result of the coherent mixing of ICD and ETMD amplitudes.

As an example, consider a hybridized dipole-allowed excited state $4p+$ from $\ful$ @$3d$ [Fig.\,1(a)] in Ar@$\ful$. The radial wavefunction of $4p+$, shown in Fig.\,1(b), results from a symmetric hybridization of free Ar and empty $\ful$ $4p$ excited states as
\begin{equation}\label{bound-hyb}
|4p+\rangle = \sqrt{\alpha}|Ar 4p\rangle + \sqrt{1-\alpha}|\ful 4p \rangle.
\end{equation} 
Based on \eq{bound-hyb}, the hybridizations in $\psi$ and ${\cal D}$ for this transition are then
\begin{equation}\label{channel-hyb}
|\psi_{@3d\rightarrow 4p+}\rangle = \sqrt{\alpha}|\psi_{@3d\rightarrow Ar4p}\rangle + \sqrt{1-\alpha}|\psi_{@3d\rightarrow \scriptsize{C_{60}} 4p}\rangle,
\end{equation}
\begin{equation}\label{ME-hyb}
{\cal D}_{@3d\rightarrow 4p+} = \sqrt{\alpha}{\cal D}_{@3d\rightarrow Ar4p} + \sqrt{1-\alpha}{\cal D}_{@3d\rightarrow \scriptsize{C_{60}} 4p}.
\end{equation}
Using Eqs.\,\ref{channel-hyb} and \ref{ME-hyb} in \eq{dc-mat-element} for the transition $@3d\rightarrow 4p+$, and assuming that the overlap between Ar $4p$ and $\ful$ $4p$ states is negligibly small, we can break up $M^{\mbox{\scriptsize d-c}}$ as
\begin{eqnarray}\label{dc-mat-element2}
{M}^{d-c}_{3s@\rightarrow kp} (E) &=& \alpha\frac{\langle\psi_{@3d\rightarrow Ar4p}|\frac{1}{|{\bf r_1}-{\bf r}_2|}
|\psi_{3s@ \rightarrow kp}(E)\rangle}{E-E_{@3d\rightarrow 4p+}} {\cal D}_{@3d\rightarrow Ar 4p}\nonumber \\
               & + & (1-\alpha) \frac{\langle\psi_{@3d\rightarrow \scriptsize{C_{60}} 4p}|\frac{1}{|{\bf r_1}-{\bf r}_2|}
|\psi_{3s@ \rightarrow kp}(E)\rangle}{E-E_{@3d\rightarrow 4p+}} {\cal D}_{@3d\rightarrow \scriptsize{C_{60}} 4p}.
\end{eqnarray}
The processes that \eq{dc-mat-element2} embodies are schematically shown in Fig.\,1. The first term denotes the release of energy from a {\em transfer} de-excitation (blue arrow) of the atomic part of the $4p+$ hybridized electron state, and the subsequent migration (purple curved arrow) of that energy to Ar knocking out a $3s$ electron (red arrow). This process is essentially an ETMD. The second term is the direct de-excitation (black arrow) of the $\ful$ part of $4p+$ followed by the regular ICD. Obviously, ETMD and ICD coherently mix to produce the ensuing resonance structure. Thus, some resonances in Fig.\,2 occur from decay rates underpinning this coherence; a detailed characterization of the structures based on Fano-fitting is forthcoming. 

To summarize, the first theoretical prediction for RICDs of fullerene innershell photoexcitations producing outershell vacancies in the caged atom is described. In fact, this generally appears to be the dominant, if not the only, single-excitation above-threshold decay mechanism through the atom's non-hybrid outer levels for any endofullerene molecule. The hybridized character of some of the excited states of the compound points to a coherence of ICD with the ETMD process. The resulting resonances are found aplenty and are quite amenable to  being probed experimentally. Although the present calculation only includes {\em participant} RICDs, where the precursor hole is filled by the excited electron itself, it is of great interest to access the influence of {\em spectator} processes; these could significantly affect the situation and certainly need study. Furthermore, with contemporary focus~\cite{dixit2013timedelay} on photoemission phase and time delay studies by interferometric metrology~\cite{kotur2015}, we hope that the current results will stimulate similar temporal spectroscopy with ICD resonances.

Finally, it is important to note that, based upon our explanation of the details of multicenter decay, this should be a strong process for any atom or molecule encaged in any fullerene, in any position, central or not. The ICD-ETMD coherence involves hybridization in the final continuum state which should be quite general -- all it requires is that both the fullerene and the trapped atom or molecule have dipole-allowed final states, continuum and quasi-discrete, of the same symmetry.
\\
\\
\\ 
{\large \bf Acknowledgment}\\
The work is supported by NSF and DOE, Basic Energy Sciences, Office of Chemical Sciences.

\end{document}